# Quantum tunneling devices incorporating two-dimensional magnetic semiconductors

Hyun Ho Kim and Adam W. Tsen, Institute for Quantum Computing and Department of Chemistry, University of Waterloo, Waterloo, Ontario N2L 3G1, Canada

**Status**

Research in two-dimensional (2D) materials has experienced rapid growth in the past few years. In particular, various layered compounds exhibiting quantum phenomena, such as superconductivity[1] and magnetism[2], have been isolated in atomically thin form, often in spite of their chemical instability. The nature of the 2D phases can be different than their bulk counterparts, making such systems attractive for fundamental studies. Owing to their high crystallinity and absence of dangling bonds, devices and heterostructures incorporating these materials may also show performance exceeding that of traditional films. In this roadmap article, we focus on a few recent developments in spin-based quantum devices utilizing the 2D magnetic semiconductor, $CrI_3$.

The 2D ferromagnetic (FM) or antiferromagnetic (AFM) compounds that have been reported thus far are: $MPS_3$ (M = Fe, Ni)[3, 4], $Cr_2Ge_2Te_6$[5], $CrX_3$ (X = I, Br, Cl)[6, 7, 8, 9, 10], $Fe_3GeTe_2$[11], and $1T\text{-}VSe_2$[12]. Since it has been rigorously proved that the 2D Heisenberg model can support neither long-range FM nor AFM order at finite temperature[13], magnetism in a monolayer should be anisotropic. Specifically, magnetism survives in monolayer $FePS_3$, $CrI_3$, $CrBr_3$, and $Fe_3GeTe_2$ owing to an out-of-plane easy axis for spin polarization. With the exception of metallic $Fe_3GeTe_2$ and $1T\text{-}VSe_2$, all exhibit semiconducting or insulating behavior: increasing resistance with decreasing temperature, and all have a magnetic transition temperature ($T_c$) below room temperature. Finally, the magnetic ions in all compounds are FM coupled within the layers, except for AFM $MPS_3$.

$CrI_3$ is particularly intriguing in that bulk and 2D forms exhibit distinct magnetic phases. Due to absence of a structural stacking transition at higher temperature, below $T_c \sim 45K$, the magnetic coupling between adjacent layers is AFM in thin samples[14], in contrast with FM coupling in the bulk. This property, combined with the Ising-type FM spin ordering within the layers, gives rise to various impressive effects when incorporated in a device heterostructures. Several groups have used few-layer $CrI_3$ as a spin-dependent barrier for quantum tunneling between graphene electrodes[8, 15, 16, 17, 18]. The application of a 2T magnetic field out-of-plane has been found to produce abrupt tunnel magnetoresistance (TMR) as large as $10^6$%. As spins in individual layers become aligned with the field, the tunnel barrier is effectively lowered, yielding an exponential rise in tunneling current (see Fig. 1a). While the magnitude of the effect decreases in thinner samples (see Fig. 1b)[15, 16, 17, 19], even in bilayer $CrI_3$ the TMR value is over twice the value than that seen from conventional EuS barriers[20]. Such devices can potentially serve as spin filters or building blocks for magnetic memory.

Using graphene as an electrostatic gate and hexagonal boron nitride (hBN) as the dielectric, two groups have been able to tune the magnetic properties of 2D $CrI_3$[21, 22, 23], thus continuing with a longstanding effort in spintronics to achieve electrical control of magnetism. In monolayer $CrI_3$, doping with hole densities of several $10^{13} cm^{-2}$ can enhance the saturation magnetization and increase $T_c$ by nearly 10%. This is not immediately expected as the exchange interaction in CrI$_3$ is not mediated by itinerant carriers. In bilayers, both doping and a pure electric field generated from two opposing gates can switch the interlayer coupling between AFM and FM ordering (see Fig. 2). Combining this with tunnelling contacts can further allow for gate-tunable TMR characteristics[24,25]. These experimental highlights demonstrate the versality and potential of 2D magnetic semiconductors for spin-based quantum devices.

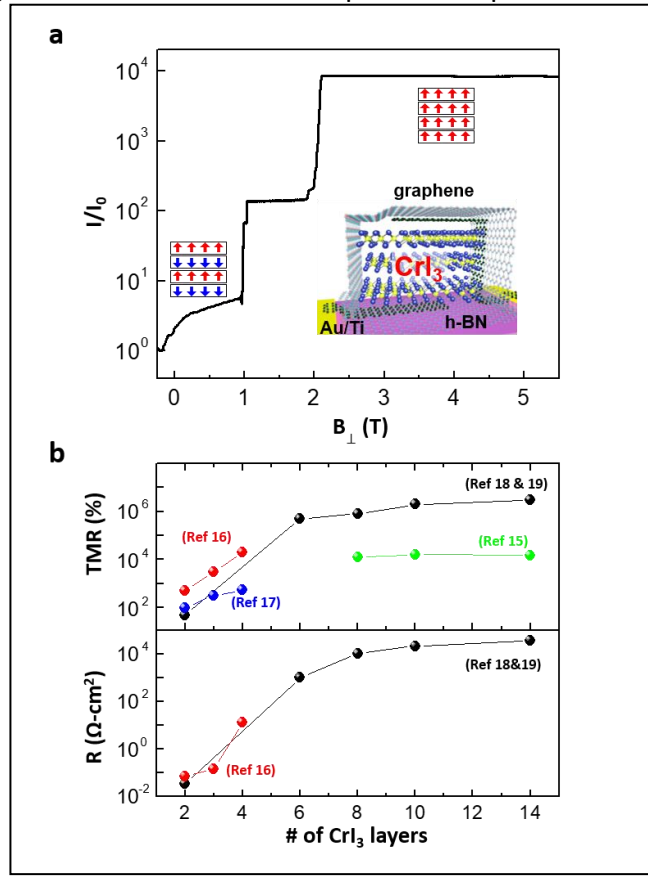

**Figure 1. Extremely large tunnel magnetoresistance (TMR) across ultrathin CrI$_3$.** (a) Change in tunneling current in few-layer CrI$_3$ device as a function of out-of-plane magnetic field at optimal voltage biasing and 1.4K. Inset shows a schematic illustration of the device. Figure reproduced with permission from: a, ref 18, American Chemical Society. (b) Thickness-dependent TMR (top) and area-normalized DC junction resistance (bottom) at optimal voltage biasing.

**Current and Future Challenges**

Despite this recent success, there are a number of challenges to be addressed in order to make such systems more technologically relevant. The most obvious limitation is that all the 2D magnetic semiconductors reported so far have a $T_c$ below room temperature, and so will not yet be suitable for practical devices. Another issue is that the conductance of tunnel devices is rather low, thus limiting their possible switching speeds. In Fig. 1b, we have plotted the DC junction resistance at the voltage bias for peak TMR as a function of CrI$_3$ thickness for all the devices reported so far in the literature normalized to their area. There is a clear trade-off in that both the resistance and TMR decrease substantially with decreasing thickness, as is expected for tunneling. Yet, even in bilayers, the resistance is ~10$^{-2}$Ω-cm$^2$, larger than that for standard magnetic tunnel junctions (~10$^{-4}$Ω-cm$^2$). Finally, for memory applications it is desirable to be able to switch between the resistive states with extremely small magnetic fields obtainable from on-chip circuit elements (~1mT). The interlayer magnetic coupling in the CrX$_3$ family yields much larger critical fields (~1T), however. While doping can be used substantially reduce this value, it appears that the interlayer AFM ground state is also destabilized[21].

**Advances in Science and Technology to Meet Challenges**

While the relative low $T_c$ is an inherent limitation of the material, there have been several reports predicting other 2D magnetic semiconductors at room temperature[26, 27]. As far as we know, they yet await experimental realization. The resistance of the tunnel junctions can be improved, in principle, by selecting metal electrodes with a lower workfunction, such as aluminum. Unfortunately, materials such as CrI$_3$ are not directly compatible with conventional fabrication procedures as they will quickly degrade in the air environment. Graphene contacts has been used as a work-around as layered heterostructures can be fully assembled in inert atmosphere. This may actually speak to a larger problem, and thus prompt the development of fabrication tools that are miniaturized in gloveboxes. Finally, the critical switching fields can be lowered by weakening the strength of the interlayer AFM coupling. This may be potentially achieved by chemically changing the interlayer spacing or even re-stacking monolayer samples with a controlled twist angle, similar to what has been demonstrated for other 2D materials.

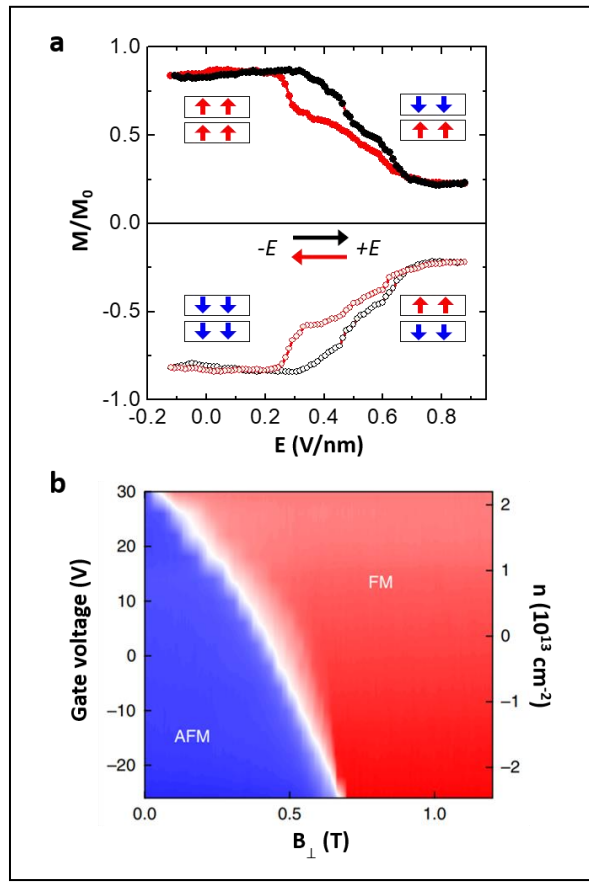

**Figure 2. Electrical control of magnetism in 2D CrI$_3$.** Effect of (a) pure electric field and (b) electrostatic doping on interlayer magnetic coupling of bilayer CrI$_3$, Figures reproduced with permission from: a, ref 23, Nature Publishing Group; b, ref 24, Nature Publishing Group.

## Concluding Remarks

Overall, the relatively young field of 2D magnetism has already led to many exciting results. The TMR physics and electrical control of magnetism may be of significant fundamental interest for the materials and device communities. Nevertheless, it remains to be seen whether the many technical hurdles can be overcome to make such systems more appealing for applications.

## Acknowledgements
We thank Shengwei Jiang for providing us with the original data for Fig. 2a.

## References

1. Saito Y, Nojima T and Iwasa Y 2016 Highly crystalline 2D superconductors. *Nat. Rev. Mater.* **2**, 16094.
2. Gong C and Zhang X 2019 Two-dimensional magnetic crystals and emergent heterostructure devices. *Science* **363**, 706.
3. Lee J-U, Lee S, Ryoo J H, Kang S, Kim T Y, Kim P, Park C-H, Park J-G and Cheong H 2016 Ising-Type Magnetic Ordering in Atomically Thin $FePS_3$. *Nano Lett.* **16**, 7433-7438.
4. Kim K, Lim S Y, Lee J-U, Lee S, Kim T Y, Park K, Jeon G S, Park C-H, Park J-G and Cheong H 2019 Suppression of magnetic ordering in XXZ-type antiferromagnetic monolayer $NiPS_3$. *Nat. Commun. 10*, 345.
5. Gong C, Li L, Li Z, Ji H, Stern A, Xia Y, Cao T, Bao W, Wang C, Wang Y, Qiu Z Q, Cava R J, Louie S G, Xia J and Zhang X 2017 Discovery of intrinsic ferromagnetism in two-dimensional van der Waals crystals. *Nature* **546**, 265-269.
*6.* Huang B, Clark G, Navarro-Moratalla E, Klein D R, Cheng R, Seyler K L, Zhong D, Schmidgall E, McGuire M A, Cobden D H, Yao W, Xiao D, Jarillo-Herrero P, and Xu X 2017 Layer-dependent ferromagnetism in a van der Waals crystal down to the monolayer limit. *Nature, 546, 270-273.*
7. Ghazaryan D, Greenaway M T, Wang Z, Guarochico-Moreira V H, Vera-Marun I J, Yin J, Liao Y, Morozov S V, Kristanovski O, Lichtenstein A I, Katsnelson M I, Withers F, Mishchenko A, Eaves L, Geim A K, Novoselov K S and Misra A 2018 Magnon-assisted tunnelling in van der Waals heterostructures based on $CrBr_3$. *Nat. Electron.* **1**, 344-349.
8. Kim H H, Yang B, Li S, Jiang S, Jin C, Tao Z, Nichols G, Sfigakis F, Zhong S, Li C, Tian S, Cory D G, Miao G-X, Shan J, Mak K F, Lei H, Sun K, Zhao L and Tsen A W 2019 Evolution of interlayer and intralayer magnetism in three atomically thin chromium trihalides. *Proc. Natl. Acad. Sci. U.S.A.* **116**, 11131-11136.
9. Klein D R, MacNeill D, Song Q, Larson D T, Fang S, Xu M, Ribeiro R A, Canfield P C, Kaxiras E, Comin R and Jarillo-Herrero P 2019 Giant enhancement of interlayer exchange in an ultrathin 2D magnet. preprint at https://arxiv.org/abs/1903.00002.
10. Cai X, Song T, Wilson N P, Clark G, He M, Zhang X, Taniguchi T, Watanabe K, Yao W, Xiao D, McGuire M A, Cobden D H and Xu X 2019 Atomically Thin $CrCl_3$: An in-Plane Layered Antiferromagnetic Insulator. *Nano Lett.* **19**, 3993-3998.
11. Deng Y, Yu Y, Song Y, Zhang J, Wang N Z, Wu Y Z, Zhu J, Wang J, Chen X H and Zhang Y 2018 Gate-tunable Room-temperature Ferromagnetism in Two-dimensional $Fe_3GeTe_2$. *Nature* **563**, 94-99.

12. Bonilla M, Kolekar S, Ma Y, Diaz H C, Kalappattil V, Das R, Eggers T, Gutierrez H R, Phan M-H and Batzill M 2018 Strong room-temperature ferromagnetism in VSe$_2$ monolayers on van der Waals substrates. *Nat. Nanotech.* **13**, 289-293.
13. Mermin N D and Wagner H 1966 Absence of Ferromagnetism or Antiferromagnetism in One- or Two-Dimensional Isotropic Heisenberg Models. Physical review letters, 17, 1133-1136.
14. Sivadas N, Okamoto S, Xu X D, Fennie C J and Xiao D 2018 Stacking-Dependent Magnetism in Bilayer CrI$_3$. *Nano Lett.* **18**, 7658-7664.
15. Wang Z, Gutiérrez-Lezama I, Ubrig N, Kroner M, Taniguchi T, Watanabe K, Imamoğlu A, Giannini E and Morpurgo A F 2018 Very Large Tunneling Magnetoresistance in Layered Magnetic Semiconductor CrI$_3$. *Nat. Commun.* **9**, 2516.
16. Song T, Cai X, Tu M W-Y, Zhang X, Huang B, Wilson N P, Seyler K L, Zhu L, Taniguchi T, Watanabe K, McGuire M A, Cobden D H, Xiao D, Yao W and Xu X 2018 Giant tunneling magnetoresistance in spin-filter van der Waals heterostructures. *Science* **360**, 1214-1218.
17. Klein D R, MacNeill D, Lado J L, Soriano D, Navarro-Moratalla E, Watanabe K, Taniguchi T, Manni S, Canfield P, Fernández-Rossier J and Jarillo-Herrero P 2018 Probing magnetism in 2D van der Waals crystalline insulators via electron tunneling. *Science* **360**, 1218-1222.
18. Kim H H, Yang B, Patel T, Sfigakis F, Li C, Tian S, Lei H and Tsen A W 2018 One million percent tunnel magnetoresistance in a magnetic van der Waals heterostructure. *Nano Lett.* **18**, 4885-4890.
19. Kim H H, Yang B, Tian S, Li C, Miao G-X, Lei H and Tsen A W 2019 Maximizing tunnel magnetoresistance across three ultrathin chromium trihalides. arXiv e-prints, https://arxiv.org/abs/1904.10476.
20. Miao G-X, Müller M and Moodera J S 2009 Magnetoresistance in Double Spin Filter Tunnel Junctions with Nonmagnetic Electrodes and its Unconventional Bias Dependence. *Phys. Rev. Lett.* **102**, 076601.
21. Jiang S, Li L, Wang Z, Mak K F and Shan J 2018 Controlling magnetism in 2D CrI$_3$ by electrostatic doping. *Nat. Nanotech.* **13**, 549-553.
22. Huang B, Clark G, Klein D R, MacNeill D, Navarro-Moratalla E, Seyler K L, Wilson N, McGuire M A, Cobden D H, Xiao D, Yao W, Jarillo-Herrero P and Xu X 2018 Electrical control of 2D magnetism in bilayer CrI$_3$. *Nat. Nanotech.* **13**, 544-548.
23. Jiang S, Shan J and Mak K F 2018 Electric-field switching of two-dimensional van der Waals magnets. *Nat. Mater.* **17**, 406-410.
24. Jiang S, Li L, Wang Z, Shan J and Mak K F 2019 Spin tunnel field-effect transistors based on two-dimensional van der Waals heterostructures. *Nat. Electron.* **2**, 159-163
25. Song T, Tu M W-Y, Carnahan C, Cai X, Taniguchi T, Watanabe K, Mcguire M A, Cobden D H, Xiao D, Yao W and Xu X 2019, Voltage Control of a van der Waals Spin-Filter Magnetic Tunnel Junction. *Nano Lett.* **19**, 915-920.
26. You J-Y, Zhang Z, Gu B and Su G 2019 Two-Dimensional Room Temperature Ferromagnetic Semiconductors with Quantum Anomalous Hall Effect. arXiv e-prints, preprint at https://arxiv.org/abs/1904.11357.
27. Fuh H-R, Chang C-R, Wang Y-K, Evans R F L, Chantrell R W and Jeng H-T 2016 Newtype single-layer magnetic semiconductor in transition-metal dichalcogenides VX$_2$ (X = S, Se and Te). *Sci. Rep.* **6**, 32625.